\documentclass[aps,prl,reprint,superscriptaddress,amsmath,amssymb,floatfix]{revtex4-2}

\usepackage[T1]{fontenc}
\usepackage{lmodern}
\usepackage{amsmath,amssymb,bm,graphicx,xcolor}
\usepackage[colorlinks=true,linkcolor=blue!45!black,citecolor=blue!45!black,urlcolor=blue!45!black]{hyperref}
\usepackage{placeins}
\DeclareMathOperator{\sn}{sn}\DeclareMathOperator{\cn}{cn}
\DeclareMathOperator{\dn}{dn}\DeclareMathOperator{\cd}{cd}
\newcommand{\B}{\boldsymbol B}\newcommand{\X}{\boldsymbol X}
\newcommand{\ii}{\mathrm i}\newcommand{\dd}{\mathrm d}\newcommand{\JJ}{\mathcal J}

\makeatletter\let\bibliography@latex\@gobble\makeatother
\DeclareSymbolFont{upgreekeuler}{U}{eur}{m}{n}
\DeclareMathSymbol{\upnu}{\mathord}{upgreekeuler}{"17}
\begin{document}
\title{Exact Nonsymmetric Vacuum Magnetic Fields with Nested Toroidal Surfaces}

\author{Wrick Sengupta}
\email{wsengupta@princeton.edu}
\affiliation{Department of Astrophysical Sciences, Princeton University, Princeton, New Jersey 08543, USA}
\author{Amitava Bhattacharjee}
\affiliation{Department of Astrophysical Sciences, Princeton University, Princeton, New Jersey 08543, USA}
\author{Taylor J. Klotz}
\affiliation{Department of Applied Mathematics, University of Colorado, Boulder, Colorado 80309-0526, USA}
\author{Andrew O. Brown}
\affiliation{Department of Astrophysical Sciences, Princeton University, Princeton, New Jersey 08543, USA}
\affiliation{Princeton Plasma Physics Laboratory, Princeton, New Jersey 08543, USA}
\author{Joshua W. Burby}
\affiliation{Department of Physics and Institute for Fusion Studies, The University of Texas at Austin, Austin, Texas 78712, USA}
\author{Vin\'icius N. Duarte}
\affiliation{Princeton Plasma Physics Laboratory, Princeton, New Jersey 08543, USA}
\affiliation{Department of Astrophysical Sciences, Princeton University, Princeton, New Jersey 08543, USA}
\author{Sanket Patil}
\affiliation{Department of Astrophysical Sciences, Princeton University, Princeton, New Jersey 08543, USA}
\author{Stefan Buller}
\affiliation{Department of Astrophysical Sciences, Princeton University, Princeton, New Jersey 08543, USA}
\date{17 October 2026}

\begin{abstract}
We give a constructive method for exact vacuum magnetic fields whose lines fill nested toroidal surfaces in ordinary
space, without continuous spatial symmetry. Complex potential theory on an algebraic curve, combined with
doubly periodic traveling waves from soliton theory, gives closed forms for the field, its potential and a surface
label. Free parameters set the rotational transform and
the magnetic shear, including exactly constant irrational transform. The same fields are exact potential flows with
toroidal stream surfaces, and give stringent tests for standard stellarator codes.
\end{abstract}
\maketitle

Vector fields whose lines fill nested tori appear across physics. They confine plasma in stellarators and tokamaks
\cite{Helander2014}, they are the stream surfaces of steady inviscid flows \cite{ArnoldKhesin}, and they organize the
knotted field lines of exact electromagnetic waves \cite{Kedia2013}. 

For a vacuum magnetic field $\B=\nabla\Phi$ with
$\nabla^2\Phi=0$, nesting is not automatic. Without a continuous symmetry, three-dimensional (3D) fields generically develop
magnetic islands and chaotic field lines \cite{CaryLittlejohn1983}. At the low pressure and small plasma current of present experiments and of reactor start-up, the confining field is essentially the vacuum field of the coils, and the quality of its flux surfaces sets the confinement volume, the edge islands and the starting point for finite-pressure equilibria. Much of the theory of stellarator vacuum fields has therefore been perturbative. Coil currents or shapes are adjusted to cancel the resonant Fourier components that open low-order islands \cite{HansonCary1984}. Quadratic-flux-minimizing and almost-invariant surfaces give integrable fields close to nonintegrable ones \cite{Dewar1994,HudsonDewar1998}, and islands in free-boundary equilibria have been removed by targeting resonant harmonics \cite{Hudson2002}. Pressure-driven currents can shrink or heal vacuum islands at finite $\beta$ \cite{Hayashi1994,Bhattacharjee1995}. These methods suppress selected islands or construct nearby integrable fields. None supplies a nonsymmetric vacuum field with guaranteed nesting through every rational surface, and so none provides an exact vacuum reference against which the approximations, and the codes built on them, can be tested. 

Grad conjectured that smooth equilibria with
nested surfaces and pressure require such a symmetry \cite{Grad1967}. Counterexamples with pressure appeared
only recently \cite{GLT2026,Landreman2026,Issan2026}. These specific constructions have nonconstant pressure and do not provide vacuum or force-free examples. Earlier exact nonsymmetric vacuum examples live on a flat toroidal quotient rather than in ordinary space \cite{WS2020}.

\begin{figure*}[t]
\includegraphics[width=\linewidth]{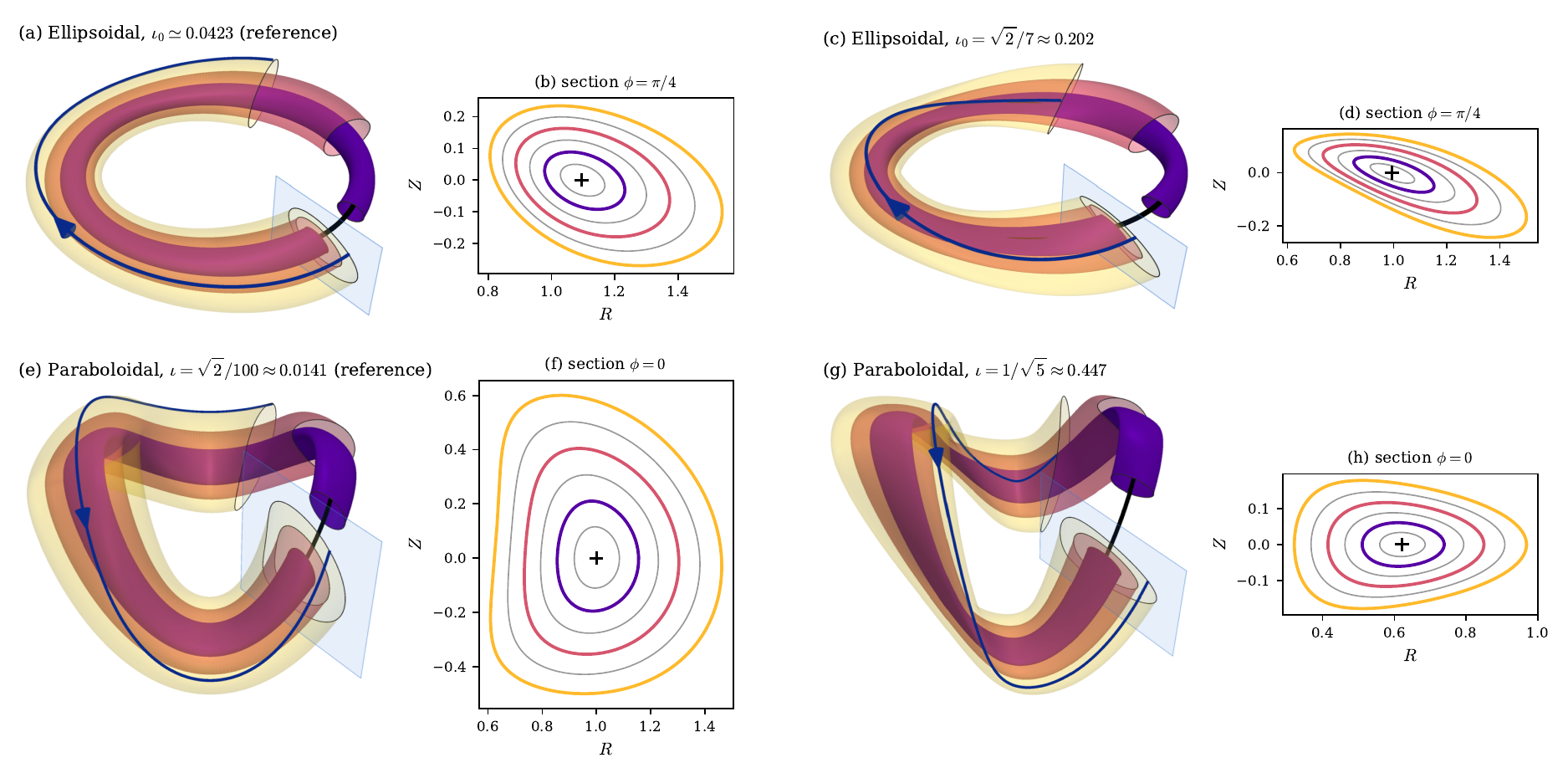}
\caption{Exact nested magnetic surfaces for the ellipsoidal (top) and paraboloidal (bottom) families, at low
rotational transform (left pair) and higher transform (right pair). In (a), (c), (e) and (g) the surfaces enclosing
the fractions $s=0.13$, $0.46$ and $1$ of the toroidal flux are cut away and partly transparent to show the nesting.
The thick black curve is the magnetic axis. The navy curve is one field line on the outer surface, drawn slightly
outside it for visibility, with an arrow along $\B$. It runs nearly parallel to the axis at low $\iota$ and winds
visibly at higher $\iota$. A faint blue plane marks the section shown to the right. Panels (b), (d), (f) and (h) show
the same three surfaces (colored) and the surfaces $s=0.04$, $0.27$ and $0.71$ (gray) in that plane, at cylindrical
azimuth $\phi=\pi/4$ for the ellipsoidal and $\phi=0$ for the paraboloidal members. Here $R$ is the cylindrical radius
and $+$ marks the axis.}
\label{fig_surfaces}
\end{figure*}

In this Letter we give a constructive method, not an isolated example, that produces exact vacuum fields in
3D configuration space whose lines fill a finite volume with nested toroidal surfaces around an axis with no
continuous rigid symmetry (Fig.~\ref{fig_surfaces}). It combines classical complex potential theory with doubly
periodic traveling waves from soliton theory, and it generates infinite families. Its free parameters set the rotational transform $\iota$, the number of poloidal turns per toroidal turn, and the magnetic shear independently, and they include members with exactly constant irrational $\iota$. Each field comes in closed form with its potential and an exact surface label. Nesting is therefore proved, not inferred from field-line
tracing or from an expansion about the axis. A companion article gives the complete construction, the proofs and the numerical
tests \cite{CompanionPRE}.

\emph{Construction.}
In two dimensions every harmonic function is the real or imaginary part of an analytic function. Whittaker found the
3D analog: harmonic functions are integrals of analytic functions over a family of complex lines
\cite{Whittaker1903}. Bergman restricted the integrand to an
algebraic curve in that space \cite{Bergman1929}. We choose curves for which the construction becomes explicit. The
curve fixes a web of confocal quadric surfaces, ellipsoidal or paraboloidal, which serve as curvilinear coordinates
$(u,v,\zeta)$ through a position map $\X(u,v,\zeta)$. The same webs carry the classical coordinates in which
Laplace's equation separates into products \cite{Eisenhart1934}. Within our paired construction, flatness fixes the
web through elliptic functions or their degenerations \cite{CompanionPRE}. Abel's theorem instead gives two of the resulting harmonic functions a
common angle $\zeta$, a toroidal angle that generally differs from the cylindrical azimuth $\phi$, so
that the potential becomes the imaginary part of a sum of an analytic
function of $u+\ii\zeta$ and one of $v+\ii\zeta$. We call the derivatives of these two functions the profiles. The
vacuum equations, a 3D generalization of the Cauchy--Riemann equations, then follow from the Cauchy--Riemann equations of the profiles and three identities of the web (Appendix~\ref{app_fields}). 

The second ingredient comes
from soliton theory. Closing the field on a torus needs profiles periodic in $\zeta$, and a regular magnetic axis needs
turning points of the profiles in $u$ and $v$, around which the level curves of the surface label close. If both
profiles are nonconstant and real on a common section of constant $\zeta$, and the label is conserved whatever the
metric of the web, then each profile $f$ obeys $f'^2=Q(f)$ with a polynomial $Q$ of degree at most four
\cite{CompanionPRE}. These are precisely the degrees for which such an equation has the Painlev\'e property, with
poles as its only movable singularities \cite{Hone2008}. The profiles' Painlev\'e property and soliton equations thus
follow from the existence of the label rather than being assumed. For our
even quartic, the solutions are Jacobi elliptic functions, with parameter $m=k^2$ and $0<k<1$ \cite{DLMF}. After
rescaling, they are periodic traveling waves of the defocusing mKdV equation
\cite{Miura1968}. These doubly periodic waves meet both requirements. Their imaginary period fits the toroidal
circuit, and the second profile is shifted by half an imaginary period, so that the two profiles have distinct turning
points that give the axis. Their addition formula, a second use of Abel's theorem, makes every field line conserve an explicit label
$\psi$. With $K=K(m)$ and $K'=K(1-m)$ the complete elliptic integrals,
$x=\kappa(u-u_0)$, $y=\kappa(v-v_0)$, $\tau_f=\kappa\zeta+\delta$, $z_1=K+x+\ii\tau_f$ and $z_2=K+y+\ii(K'+\tau_f)$, the potential is
\begin{align}\label{prl_potential}
 \Phi&=\kappa^{-1}\operatorname{Im}[F(z_1)+F(z_2)],\nonumber\\
 F(z)&=k^{-1}\log[\dn(z|m)-k\cn(z|m)].
\end{align}
The parameters are the axis location $(u_0,v_0)$, the scale $\kappa$, the phase $\delta$ and $k$, together with the
shape parameter $m_g$ of the confocal ellipsoids (Appendix~\ref{app_fields}). Writing $L$ for the period of $\zeta$, the period condition $\kappa L=4NK'$
with integer $N\ge1$ makes the field single valued, while $\Phi$ gains the constant $4\pi N/(\kappa k)$ per toroidal
circuit. This is the circulation of a curl-free field around the hole of the torus. With
$U=\cd(x|m)$, $V=[k\cd(y|m)]^{-1}$ and $(s_f,c_f,\Lambda_f)=(\sn\tau_f,\cn\tau_f,\dn\tau_f)|_{1-m}$, the surface label is
\begin{equation}\label{prl_label}
 \psi^2=\frac{\Lambda_f^2\,(U-V)^2}{(c_f^2+mU^2s_f^2)(c_f^2+mV^2s_f^2)},\qquad\psi>0.
\end{equation}
The explicit maps and the Cartesian field for both geometries are given in Appendix~\ref{app_fields}. Landreman's counterexamples to Grad's conjecture also carry soliton structure. Along each field line, parametrized by $t$ with $\dd\X/\dd t=\B$, a constant multiple of a signed field component is a Jacobi elliptic traveling wave of the focusing mKdV equation \cite{Landreman2026,CompanionPRE}. 

\begin{figure*}[t]
\includegraphics[width=0.388\linewidth]{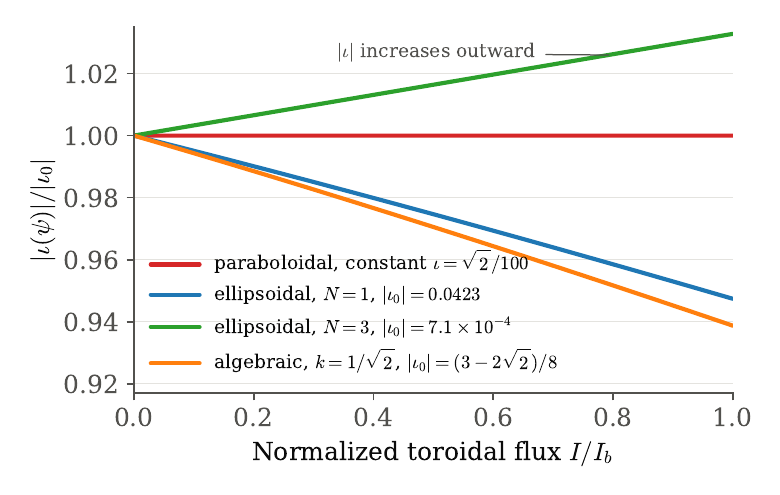}\hfill
\includegraphics[width=0.597\linewidth]{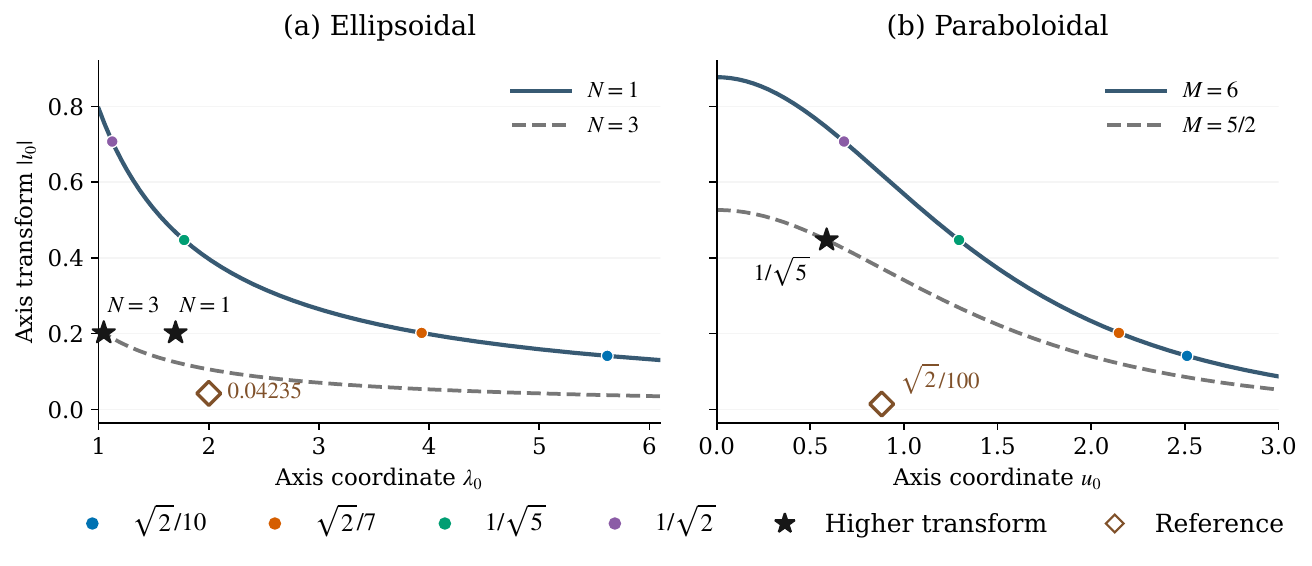}
\caption{Independent control of transform and shear. Left: transform magnitude divided by its axis value against
normalized toroidal flux for four members. The paraboloidal member has exactly constant irrational transform. The two
ellipsoidal members with $N=1$ and $N=3$ profile repetitions have decreasing and increasing transform magnitude, respectively, and the algebraic member
has $\iota_0=(3-2\sqrt2)/8$. Right: axis transform $|\iota_0|$ against the axis coordinate for the two families. Solid
curves are one parameter family each, and dashed curves a second. Colored points mark exact targets
$\sqrt2/10$, $\sqrt2/7$, $1/\sqrt5$ and $1/\sqrt2$, stars mark the members of Figs.~\ref{fig_surfaces}, \ref{fig_codes} and \ref{fig_orbit}, labeled by $N$
for the ellipsoids, and diamonds the reference members.}
\label{fig_control}
\end{figure*}

\emph{Nested surfaces.}
In each regular domain the field is analytic, nonvanishing and satisfies
\begin{equation}\label{prl_structure}
 \B=\nabla\Phi,\qquad \nabla^2\Phi=0,\qquad \B\cdot\nabla\psi=0,\qquad \B\cdot\nabla\zeta>0.
\end{equation}
The last inequality makes every line advance around the torus. The label has its strict minimum $\psi_0=k^{-1}-1$ on
the closed curve $(u,v)=(u_0,v_0)$, which is the magnetic axis, and increases strictly along every outward coordinate
ray. Where the position map is one to one, each level $\psi=c>\psi_0$ is therefore a smooth torus, the levels are
nested, and they shrink smoothly onto the axis \cite{CompanionPRE}. This proves nesting through the whole domain,
including every rational surface. Rational surfaces carry closed field lines with no islands. The axes are
noncircular and isolated, which excludes any continuous rigid symmetry.

\emph{Control of transform and shear.}
Let $2\pi I(\psi)$ and $2\pi J(\psi)$ be the toroidal and poloidal magnetic fluxes inside the surface $\psi$. With primes
denoting $\dd/\dd\psi$,
\begin{equation}\label{prl_flux}
 \iota=J'(\psi)/I'(\psi),\qquad \dd\iota/\dd I=\iota'(\psi)/I'(\psi),
\end{equation}
and both fluxes are explicit integrals of the closed-form field, with no fitting of field-line
returns. Near the axis the quadratic part of the flux action $I$ is the Ermakov--Lewis invariant \cite{Lewis1967,Bernardin1985} of the linearized
flow in a periodic canonical gauge, and the exact action $I(\psi)$ continues it to finite radius \cite{CompanionPRE}. In the paraboloidal family the two flux derivatives are proportional on every surface, so $\iota$ is
exactly constant and irrational for suitable parameters, for example $\iota=\sqrt2/100$ or $1/\sqrt5\approx0.447$. The ellipsoidal family has nonzero shear, and $|\iota|$ can rise or fall outward (Fig.~\ref{fig_control}, left).
The axis transform follows from closed-form residue sums (Appendix~\ref{app_fields}) and is moved continuously by the axis location
(Fig.~\ref{fig_control}, right). Exact members with $\iota_0=\sqrt2/7\approx0.202$ include one whose transform falls by
$27\%$ across its domain and one whose transform rises by $2\%$. A Gardner deformation of the profiles, which turns the mKdV waves into
traveling wave solutions of the Gardner equation, changes the axis shear while preserving $\iota_0$. Other parameters vary the mirror structure. Nested tori around the axis persist
over continuous parameter ranges, and only the extent of the nested region is found member by member (Appendix~\ref{app_domains}).

\begin{figure*}[t]
\includegraphics[width=\linewidth]{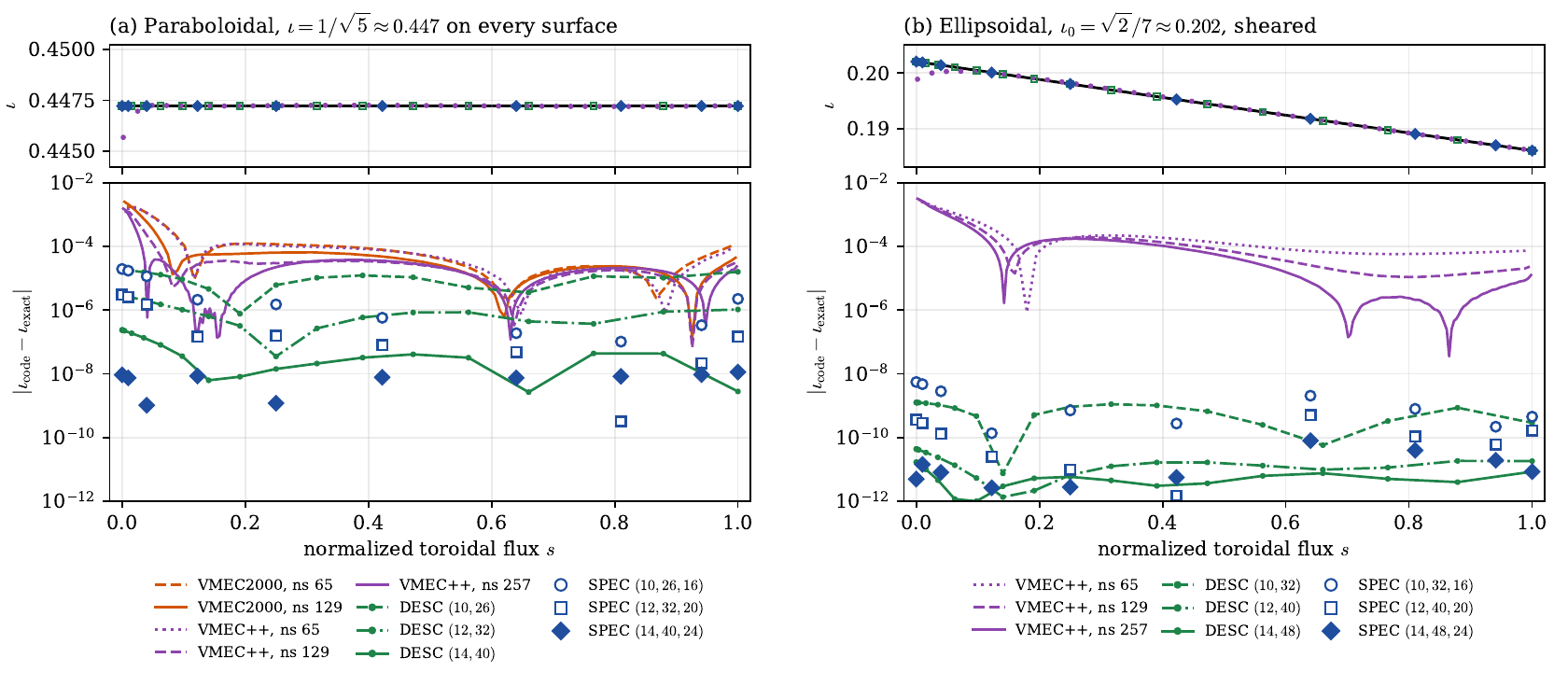}
\caption{Rotational transform error of four codes against normalized toroidal flux $s$, for (a) the paraboloidal member with
constant $\iota=1/\sqrt5$ and (b) the ellipsoidal member of Fig.~\ref{fig_surfaces}(c,d) with $\iota_0=\sqrt2/7$. Upper strips
show the exact transform (black) and the finest run of each code.
VMEC2000 did not converge in (b) and is omitted. The rise toward $s=0$ is the known VMEC inaccuracy near the magnetic axis \cite{Panici2023DESC}.
Legends give $n_s$ for VMEC, $(M,N,L_{\rm rad})$ for SPEC and $(L,N)$ for DESC.}
\label{fig_codes}
\end{figure*}

\begin{figure*}[t]
\includegraphics[width=\linewidth]{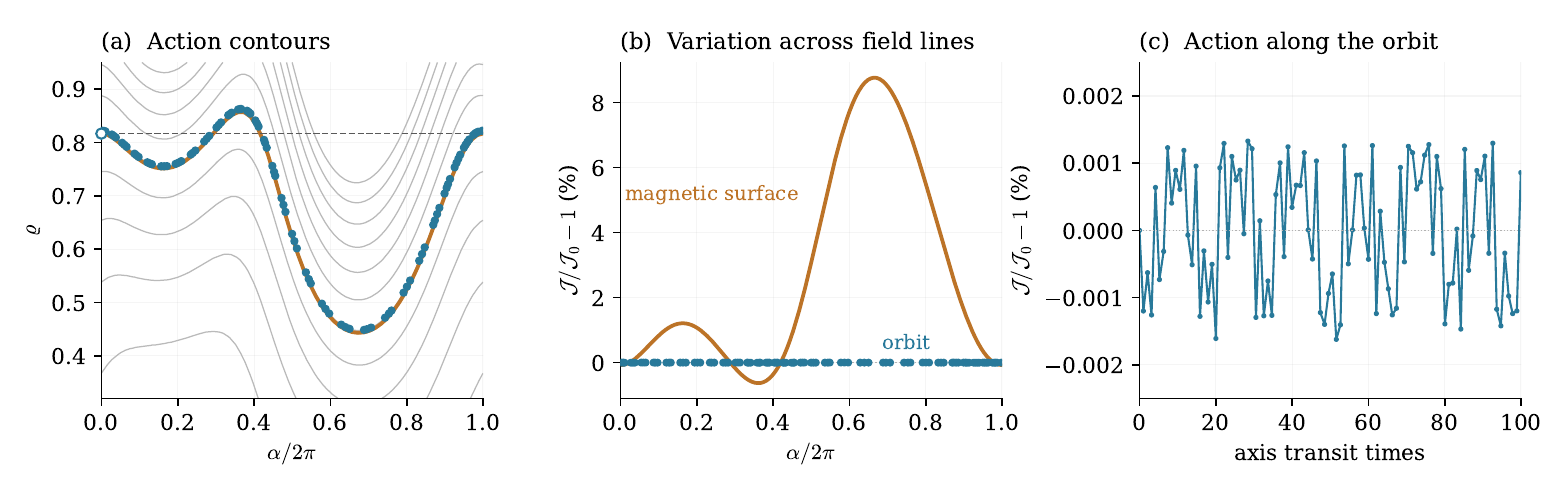}
\caption{A trapped guiding-center orbit in the ellipsoidal member with $N=3$ and $\iota_0=\sqrt2/7$, at mirror-field
depth fraction $0.35$ and $\rho_L/a=0.005$. (a) Action contours (gray) against the field-line label $\alpha$ and
the label radius $\varrho$, the square root of the normalized surface label, so that magnetic surfaces are horizontal
lines. The closed contour through the launch (orange), 96 turning points of the orbit mapped back along field lines
(blue) and the launch magnetic surface (dashed) are shown. (b) Action relative to $\JJ_0$, its value at the first turning point of the orbit, around the launch magnetic surface
(orange) and along the orbit (blue). (c) Action along the orbit against time in axis transit times. The launch is not
a turning point. The guiding center drifts about one Larmor radius before its first bounce, so the action of the launch
field line is $0.08\%$ below $\JJ_0$.}
\label{fig_orbit}
\end{figure*}

\emph{Stringent tests for equilibrium codes.}
The exact outer surface, the toroidal flux, zero pressure and zero toroidal current are sufficient inputs for an
equilibrium code. The interior field, the axis and the transform are then predictions that the exact solution checks
directly. Figure~\ref{fig_codes} shows four codes \cite{Hirshman1983VMEC,VMECpp2026,Hudson2012SPEC,Panici2023DESC} on
the paraboloidal member with $\iota=1/\sqrt5$ and on the sheared ellipsoidal member of Fig.~\ref{fig_surfaces}(c,d).
Away from the axis the VMEC transform error falls with radial resolution to a few times $10^{-5}$, but next
to the axis it stays between $10^{-3}$ and $3\times10^{-3}$ at every tested resolution, the known inaccuracy of
VMEC there \cite{Panici2023DESC}. VMEC2000 did not converge on the ellipsoidal member. SPEC, given the exact axis as its coordinate axis, reaches
errors of order $10^{-8}$ on the paraboloidal and $10^{-10}$ on the ellipsoidal member. DESC, which starts from the exact surfaces, reaches $2\times10^{-7}$ on the paraboloidal and
$2\times10^{-11}$ on the ellipsoidal member. The exact solutions measure the accuracy of the computed field directly. These comparisons illustrate how demanding the fields are as tests.

\emph{Particle orbits.}
Nested magnetic surfaces confine field lines, not particles, which drift across the surfaces. For trapped particles
at fixed energy and magnetic moment, bounce averaged drifts follow contours of the action
$\JJ=\int_{l_-}^{l_+}(B_m-|\B|)^{1/2}\,\dd l$, where $B_m$ is the mirror field and $l$ is arclength between the
mirror points. Because the field and its derivatives are explicit, $\JJ$ can be evaluated to quadrature precision. In
the ellipsoidal member with $\iota_0=\sqrt2/7$ and $N=3$, one trapped class has a closed action contour that crosses magnetic surfaces
(Fig.~\ref{fig_orbit}), an iso-action class in the sense of Ref.~\cite{SenguptaIsoaction2026}, which requires
only that the drift surface close. Around the launch surface the action varies by $9.4\%$, yet along the full guiding-center
trajectory, over three drift circuits, it varies by only $0.003\%$. This is a single finite-time
orbit. It shows how the exact fields isolate questions of particle confinement from those of field-line confinement, but it does not establish omnigeneity.

\emph{Broader implications.}
The same fields are exact solutions of steady inviscid fluid flow. With velocity $\mathbf v=\B$ and pressure
$p=C-|\mathbf v|^2/2$ at unit density, they satisfy the incompressible Euler equations with zero vorticity
\cite{ArnoldKhesin}. They are therefore three-dimensional potential flows, the basic model of inviscid aerodynamics, with circulation
through the hole of the torus, nested toroidal stream surfaces and no symmetry. In twistor terms, Abel's theorem on a compact algebraic curve in the space of oriented lines gives the common angle,
and the same theorem on the elliptic curve $y^2=Q(x)$ of the soliton profiles gives the surface label
\cite{PenroseTwistor,Dunajski2003,CompanionPRE}.  A KdV approach to quasisymmetric magnetic fields selects related KdV and Gardner reductions by requiring the
Painlev\'e property. Deriving that property from quasisymmetry alone remains open \cite{Sengupta2026}.
A related complex construction underlies exact electromagnetic fields whose lines form torus knots on nested tori \cite{Kedia2013}. Our fields
are magnetostatic, curl-free analogs, with closed torus-knot field lines on every rational surface and dense lines on
every irrational one.  Finally, for Hamiltonian dynamics they are explicit
integrable systems without symmetry. 

In summary, combining complex potential theory with doubly periodic traveling waves from soliton theory yields
families of exact vacuum magnetic fields with nested toroidal surfaces in ordinary space, no continuous rigid
symmetry, closed-form potentials and labels, and independently controlled transform and shear.  Finding external currents that produce a chosen member, and the robustness of
the surfaces to perturbations, are design questions for future work.

\FloatBarrier
\begin{acknowledgments}
Under the authors' direction, OpenAI Codex (recorded model identifiers gpt-6-astra, gpt-6-sol and gpt-6.1-sol) and
Anthropic Claude Code (Claude Opus 5.5 and Claude Fable 5.1) assisted with analytical derivations, symbolic and
numerical calculations, equilibrium code comparisons, orbit computations, assistance in figure preparation, literature searches and manuscript
drafting. The authors set the research questions, selected the examples and reviewed intermediate results and
drafts. Checks included analytical identities, symbolic calculations, high-precision evaluations, convergence tests
and comparisons between separate agent sessions, as described in the paper. The authors retain responsibility for
the scientific claims and the final manuscript. This research was supported by the Simons Foundation/SFARI (Grant No.\ 560651, AB) and the Department of Energy
(Award No.\ DE-SC0024548).
\end{acknowledgments}

\appendix
\setcounter{secnumdepth}{2}
\section{Explicit fields}\label{app_fields}
With $b=\operatorname{Im}\sn(z_1|m)$, $e=\operatorname{Im}\sn(z_2|m)$ and $A=\operatorname{Re}[\sn(z_1|m)+\sn(z_2|m)]$,
the derivatives of the potential are $\Phi_u=b$, $\Phi_v=e$ and
$\Phi_\zeta=A$. Both geometries share the Euclidean metric
\begin{equation}\label{em_metric}
 \dd\X^2=P^2(\dd u^2+\dd v^2)+2PE\,\dd u\,\dd v+D\,\dd\zeta^2,
\end{equation}
with $D=P^2-E^2>0$ and $P>0$. Writing $h=E/P$, the Cartesian field at $\X(u,v,\zeta)$ is
\begin{equation}\label{em_field}
 \B=\frac{(b-he)\X_u+(e-hb)\X_v+A\X_\zeta}{D},
\end{equation}
so that $\B\cdot\nabla\zeta=A/D>0$.

\subsection{Paraboloidal geometry}
Here $L=2\pi$ and
\begin{equation}\label{em_pmap}
 \X=\begin{pmatrix}\sinh u\cos\zeta-\sinh v\sin\zeta\\ \sinh u\sin\zeta-\sinh v\cos\zeta\\
 \sinh u\sinh v+\tfrac12\sin2\zeta\end{pmatrix},
\end{equation}
with $P=\cosh u\cosh v$ and $E=\sinh u\sinh v-\sin2\zeta$. The member with $\iota=1/\sqrt5$ has $K'/K=5/2$, $u_0=\operatorname{arcosh}[\sqrt5\,\operatorname{sech}(2\pi/5)]$,
$v_0=\delta=0$ and $\kappa=2K'/\pi$, and every surface has $\iota=\operatorname{sech}(\pi K/K')/\cosh u_0$.

\subsection{Ellipsoidal geometry}
With the shape parameter $0<m_g<1$, put $R=(u+v)/4$, $S=(u-v)/4$ and
$\tau_g=\zeta/2+K(1-m_g)/2$, so that $L=8K(1-m_g)$. Then
\begin{equation}\label{em_emap}
 \X=\frac{1}{\sn R}\begin{pmatrix}\cn R\,\sn\tau_g\,\dn S\\ \dn R\,\cn\tau_g\,\cn S\\ \dn\tau_g\,\sn S\end{pmatrix},
\end{equation}
where functions of $R$ and $S$ have parameter $m_g$ and those of $\tau_g$ have parameter $1-m_g$. With
$\lambda=\sn^{-2}R$, $\upnu=m_g\sn^2S$ and $\mu=1-(1-m_g)\sn^2\tau_g$, one has $P=(\lambda-\upnu)/4$ and
$E=(\lambda+\upnu-2\mu)/4$. The ellipsoidal reference member has $k=m_g=1/2$, $u_0=v_0=2F(\pi/4|1/2)$,
$\delta=0$ and $\kappa=K(3/4)/[2K(1/2)]$, where $F$ is the incomplete elliptic integral. At fixed $u,v$ the
cylindrical azimuth $\phi$ of $\X$ changes monotonically, by $2\pi$ per circuit of $\zeta$ in the paraboloidal map,
where $\phi=\zeta$ on $v=0$, and by $-2\pi$ in the ellipsoidal map.

\subsection{Vacuum equations}
The metric \eqref{em_metric} gives $\B\cdot\X_u=b$, $\B\cdot\X_v=e$ and $\B\cdot\X_\zeta=A$.
Since $\sn(z_1|m)$ and $\sn(z_2|m)$ are analytic on the regular domain, their Cauchy--Riemann equations give
\begin{gather}
 b_v=e_u=0,\qquad b_\zeta=A_u,\qquad e_\zeta=A_v,\nonumber\\
 b_u+e_v+A_\zeta=0.\label{em_cr}
\end{gather}
The components of $\nabla\times\B$ along $\X_u$, $\X_v$ and $\X_\zeta$ are $(A_v-e_\zeta)/J$,
$(b_\zeta-A_u)/J$ and $(e_u-b_v)/J$, with $J=\det(\X_u,\X_v,\X_\zeta)$, so the first line gives
$\nabla\times\B=0$. With the volume element $PD$,
Eqs.~\eqref{em_field} and \eqref{em_cr} give
\begin{align}
 PD\,\nabla\cdot\B&=\partial_u(Pb-Ee)+\partial_v(Pe-Eb)+\partial_\zeta(PA)\nonumber\\
 &=b(P_u-E_v)+e(P_v-E_u)+AP_\zeta,\label{em_div}
\end{align}
which vanishes because $P_u=E_v$, $P_v=E_u$ and $P_\zeta=0$ in both geometries. In the paraboloidal one
$P_u=E_v=\sinh u\cosh v$ and $P_v=E_u=\cosh u\sinh v$. In the ellipsoidal one $\lambda$, $\upnu$ and $\mu$ depend only
on $R$, $S$ and $\tau_g$, so $P_u=E_v=(\lambda_R-\upnu_S)/16$ and $P_v=E_u=(\lambda_R+\upnu_S)/16$.

\subsection{Fluxes}
Write $\det D\X=\sigma PD$ with $\sigma=\pm1$. On the section $\tau_f=0$ the label reduces to
$\psi=V-U$ and $\Phi_\zeta=U+V$, so the toroidal flux $2\pi I$ inside the surface $\psi$ is
$I(\psi)=(2\pi)^{-1}\int_{\mathcal D_\psi}P\,(U+V)\,\dd u\,\dd v$, where $\mathcal D_\psi$ is the connected component of $V-U\le\psi$ that contains the axis point $(u_0,v_0)$. The
poloidal flux $2\pi J$ crosses the ribbon $v=v_0$, $u_0\le u\le u_+(\psi,\zeta)$, where $u_+$ meets the level $\psi$,
so $J(\psi)=-(\sigma/2\pi)\int_0^L\int_{u_0}^{u_+}E\,\Phi_u\,\dd u\,\dd\zeta$.

\subsection{Axis transform}
Let $K_g=K(m_g)$ and $K_t=K(1-m_g)$, with nome ratios $M=K'/K$ and $G=K_t/K_g$ such that
$H=NM/G$ is an odd integer. Then $\kappa=NK'/(2K_t)$. For a planar axis, $v_0=u_0$ and $\delta=(N-1)K'/2$, the axis
transform is $\iota_0=\mathcal S_N/\lambda_0$ with $\lambda_0=\sn^{-2}(u_0/2|m_g)$ and the finite sum
\begin{equation}\label{em_SN}
 \mathcal S_N=\sum_{j=0}^{H-1}\sum_{p=0}^{2N-1}(-1)^{j+p+1}\dn^2(w_{jp}|1-m_g),
\end{equation}
with $w_{jp}=(p+1)K_t/N+\ii(2j+1)K_g/H$. Its single pole entry, $j=(H-1)/2$ and $p=2N-1$, is replaced by its regular value $(1+m_g)/3+2\kappa^2(1-6k+k^2)/3$
\cite{CompanionPRE}. Any $0<\iota_0<\mathcal S_N$ therefore fixes a closed-form member. The member with
$\iota_0=\sqrt2/7$ in Figs.~\ref{fig_surfaces} and \ref{fig_codes} has $N=1$ and $M=G=2$, so $H=1$, $k=3-2\sqrt2$,
$m_g=k^2$ and $\kappa=1/2$. Then $\dn(K_t+\ii K_g|1-m_g)=0$ and $1-6k+k^2=0$ give $\mathcal S_1=(1+m_g)/3=6-4\sqrt2$,
and $\lambda_0=21\sqrt2-28$ with $\delta=0$ gives $\iota_0=\sqrt2/7$ exactly.

\subsection{Shear at fixed axis transform}
For the reference ellipsoid, the Gardner deformation replaces the profiles
$p_1=\sn(z_1|m)$ and $p_2=\sn(z_2|m)$ by $p_1/(1+\epsilon p_1)$ and $p_2/(1-\epsilon p_2)$. The deformed field remains harmonic and has an
exact label. On a common finite tube for $|\epsilon|\le1/10$, its axis transform is unchanged and its axis shear is
multiplied by $1+2\epsilon^2$ \cite{CompanionPRE}.

\section{Domains and checks}\label{app_domains}
The restrictions $|x|,|y|<K$ exclude the poles of the profiles. The maps are one to one and regular for
$0<R<K(m_g)$, $|S|<K(m_g)$ in the ellipsoidal geometry and for $u>|v|$ in the paraboloidal geometry. A level is a
torus of the family when it stays inside both sets of restrictions over a full circuit, which is certified for each
member \cite{CompanionPRE}. The agents identified in the Acknowledgments assisted in developing and running the calculation scripts. The
analytical checks, numerical tolerances and comparisons used to assess the results are given here and in the
main text. Independent implementations of Eqs.~\eqref{prl_potential}, \eqref{prl_label} and
\eqref{em_metric} to \eqref{em_emap} at 30 to 40 digits reproduce $\B\cdot\nabla\psi=0$, $\psi=\psi_0$ on the axis
and the stated circulation of $\Phi$ to residuals below $10^{-30}$.

\end{document}